                                 \documentclass[12pt]{article}
                                 \usepackage{aaspp4}

\def\feka{Fe~K$\alpha$}
\def\fekb{Fe~K$\beta$}
\def\nika{Ni~K$\alpha$}
\def\fekba{\fekb/\feka}

\def\sax{{\it BeppoSAX}}
\def\etal{et al. }

\begin{document}

\title{Central Elemental Abundance Ratios in the Perseus Cluster: 
 Resonant Scattering or SN Ia Enrichment?}

\author{Renato A. Dupke}
\affil{Department of  Astronomy, University of Michigan,
   Ann Arbor, MI~48109-1090}
\author{Keith A. Arnaud}
\affil{NASA/GSFC, Laboratory for High Energy Astrophysics, Greenbelt, 
MD~20771}
\affil{Department of Astronomy, University of Maryland, College Park, MD~20742}

%===============================================================================
\begin{abstract} 
We have determined elemental abundance ratios in the core of the
Perseus cluster for several elements. These ratios indicate a central
dominance of SN Ia ejecta similar to that found for A496, A2199 and
A3571 (Dupke \& White 2000a). Simultaneous analysis of {\sl ASCA} spectra from
SIS1, GIS 2\&3 shows that the ratio of Ni to Fe abundances is $ \sim
3.4 \pm 1.1 $ times Solar within the central 4$^\prime$. This ratio is
consistent with (and more precise than) that observed in other
clusters whose central regions are dominated by SN Ia ejecta.  Such a
large Ni over-abundance is predicted by ``Convective Deflagration'' explosion models for SN
Ia such as ``W7'' but is inconsistent with delayed detonation
models. We note that with current instrumentation the \nika\ line is
confused with \fekb\ and that the Ni over-abundance we observe has
been interpreted by others as an anomalously large ratio of \fekb\ to
\feka\ caused by resonant scattering in the \feka\ line. We argue that
a central enhancement of SN Ia ejecta and hence a high ratio of Ni to
Fe is naturally explained by scenarios that include the generation of
chemical gradients by suppressed SN Ia winds or ram-pressure
stripping of cluster galaxies. It is not necessary to suppose that
the intracluster gas is optically thick to resonant scattering of the \feka\ line.
\end{abstract}
 
                                \clearpage
%===============================================================================
                                \section{
Introduction
                                }
 
The existence of heavy elements in the intracluster medium (ICM) has
been known since the discovery of the $\sim 6.7$ keV \feka\ line in
the Perseus cluster (Mitchell et al. 1976) and indicates that part of
the ICM is not primordial, i.e. was processed in stars and injected
into the ICM. If the emitting plasma is assumed to be optically thin
then the iron line flux can be converted into an elemental
abundance. When this is done the inferred ICM iron abundances are
typically in the range 0.3--0.4 Solar (Mushotzky \& Loewenstein 1997). The
measurements of abundances of iron and other elements are key data
used in constraining models of the formation and evolution of the
ICM. Therefore, it is important to check that line fluxes can be
unambiguously converted to elemental abundances.

A key assumption is that the plasma is optically thin. However,
Gilfanov, Sunyaev \& Churazov (1987) pointed out that this may not
always be true for the \feka\ line. Resonant scattering (the
absorption and immediate re-emission of an \feka\ line photon by an Fe
ion) could be significant in the cores of clusters. If so, the line
emission from the cluster core would be reduced because of photons
being scattered out of the line of sight and the measured Fe
abundance would be an underestimate of the true intracluster gas
abundance.  The simplest test for resonant scattering in the \feka\
line is to compare the line flux with that from the \fekb\ line. The
optical depth in the \fekb\ line is $\sim 20\%$ of the \feka\
 so that if the \fekba\ ratio is anomalously high then resonant
scattering must be taken into account.

Such anomalous ratios have been observed with low significance in data
from the HEAO 1 (Mitchell \& Mushotzky 1980) and {\sl Tenma\ }
(Okumura \etal 1988) satellites. Recently, these results have been
placed on a much firmer basis using observations of several clusters
using {\sl ASCA} (Akimoto \etal 1997; Yamashita \etal 1999) and of the Perseus
cluster using \sax\ (Molendi \etal 1998). The best data are from
Molendi \etal\ who show that the \fekba\ ratio is anomalously
high for the inner 6 $^\prime$ of the Perseus cluster. Outside that
radius the ratio decreases to the expected value for an optically 
thin plasma. They deduce from this that the Fe abundance is underestimated 
by about a factor of 2 in the center of the Perseus cluster. 

However, at the resolution of current X-ray spectrometers the \fekb\
line is confused with the \nika\ line so that a measurement of the \fekba\
ratio requires an assumed flux in the \nika\ line. The results
reported above assumed that the ratio of Ni and Fe abundances is
similar to the Solar value (0.038 by number - photospheric value;
Anders \& Grevesse 1989). To make the \fekba\ ratio consistent with
the optically thin case requires a Ni/Fe abundance ratio several times
that of Solar. To see whether this is reasonable it is necessary to
consider how heavy elements arrived in the ICM.

$Einstein$ FPCS spectroscopy (Canizares et al.\ 1982) and more recent
{\sl ASCA} spectroscopy (Mushotzky \& Loewenstein 1997; Mushotzky et
al.\ 1996) suggest that global intracluster metal abundances are
consistent with ejecta from Type II supernovae (SN II). This implies
that the bulk of the heavy elements were injected into the ICM by
protogalactic winds. This is further supported by analysis of ICM energetics (White 1991).
 Since SN II do not produce much Ni we do not
expect Ni/Fe abundance ratios large enough to restore the \fekba\
ratio to the optically thin value. However, there are recent indications that SN Ia ejecta can 
contribute with a significant fraction (up to 50\%) of the ICM metals
(Ishimaru \& Arimoto 1997; Fukazawa et al.\ 1998; Nagataki \& Sato 1998, 
Finoguenov \& Ponman 1999, Allen et al. 2000).

Clues to metal enrichment mechanisms in clusters can be derived from the analysis 
of clusters with central abundance enhancements (abundance gradients).
Analysis of spatially resolved X-ray spectra has shown that in several clusters 
the metal abundance is centrally enhanced (Koyama, Takano \& Tawara 1991; 
White et al.\ 1994; Fukazawa et al.\ 1994; Ulmer et al.\ 1987; Ponman et al.\ 1990;
Kowalski et al.\ 1993; Arnaud, et al.\ 1994; Ikebe et al.\ 1997; Ezawa et al.\ 1997; 
Xu et al.\ 1997; Pislar et al.\ 1997; Dupke 1998; Dupke \& White 2000a,b; 
Irwin \& Bregman 1999; Finoguenov, David \& Ponman 1999; White 1999; Allen et al. 1999). 
There are several mechanisms which may cause central abundance enhancements in 
the ICM including: 1) the radial profile of
intracluster gas density is shallower than that of metal-injecting
galaxies, i.e., if metals trace the galaxies that injected them into the 
ICM and these galaxies are more concentrated in the center than the 
intracluster gas one would expect to find a general abundance 
enhancement towards 
the clusters' central regions; 2) mass loss from
the stars in central dominant galaxies may accumulate near the cluster
center; 3) ram-pressure stripping of the metal-rich gas in cluster
galaxies by intracluster gas is more effective at the center, where
the intracluster gas density is highest; 4) secular 
SN Ia winds in central dominant galaxies which are partially suppressed 
due to the galaxy location at the bottom of the cluster gravitational
potential, where the intracluster gas density is highest (Dupke \&
White 2000b). Some of the above mentioned mechanisms could deposit substantial 
amounts of SN Ia enriched material near the clusters' center, therefore creating a 
``chemical gradient''. Since SNe Ia
 and II have different elemental mass yields, analysis of the elemental abundance ratios
provides information on the fractional contribution of gas enriched by SN Ia and II. 
The analysis of abundance ratios 
in the central regions of clusters having abundance gradients indeed indicates 
a dominance of SN Ia ejecta (Dupke \& White 2000a,b see below).

As mentioned above, SN Ia metal injection mechanisms such as
ram-pressure stripping or post-protogalactic suppressed SN Ia winds
may contribute significantly to ICM enrichment. SN Ia can generate
large amounts of Ni. ``Convective Deflagration'' models such as W7
(Nomoto, Thieleman, \& Yokoi 1984; Thieleman, Nomoto \& Yokoi 1986)
predict a Ni/Fe number abundance ratio $\sim 5$ times
Solar. Alternative slow flame-speed (delayed detonation) models
produce a lower Ni/Fe abundance ratio closer to that from SN II
($\approx$ 1.6). Recently, Dupke \& White (2000a) and Dupke (1998) have
found evidence that in the central regions of A496, A2199, and A3571
the ICM is enriched predominantly by SN Ia ejecta. In these objects
the ratio of Fe abundance to that of $\alpha$-process elements
increases with decreasing radius from the cluster center. These
results suggest that, in clusters for which the abundance profile is
centrally enhanced, SN Ia ejecta provide most of the total iron mass
in the central regions (see also Allen et al. 1999).  In the outer
regions the SN Ia contribution is smaller than in the cluster's center
but not negligible (e.g. Nagataki \& Sato 1998). In A496 the excess
iron that generates the central abundance gradient is fully accounted
for by the excess SN Ia ejecta that produce a chemical gradient (Dupke
\& White 2000b). In all three clusters mentioned above the Ni/Fe
abundance ratio is consistent with that predicted by the SN Ia
``Convective Deflagration'' W7 model, i.e., the Ni/Fe ratio is several
times Solar (assuming that the \fekba\ ratio takes its optically thin
value). The analysis of A1060 (the only cluster with flat abundance
distribution in Dupke's sample (1998)) shows no significant abundance
ratio trend in its radial distribution (including the Ni/Fe ratio).
However, since the number of clusters with detailed measured abundance
ratio distributions is relatively small, it is not clear whether this
``chemical gradient'' is always true for clusters with central abundance
gradients.

This work extends the analysis of Dupke \& White (2000a) to the
Perseus cluster. This cluster is closer and brighter than A496, A2199,
and A3571 so we can determine elemental abundance ratios with
higher precision. The Fe abundance is known to be centrally enhanced
in Perseus (Ulmer \etal 1987; Ponman \etal 1990; Kowalski et al.\ 1993; Arnaud
\etal 1994; Molendi \etal 1998). If gradients in abundance and abundance ratios 
 are related there should be a significant
contribution from SN Ia ejecta in the regions where the abundance is
enhanced. We argue below that the central regions of the Perseus
cluster do show evidence for SN Ia ejecta dominance, that the Ni/Fe abundance
ratio is that predicted by the W7 
models, and hence there is no evidence for resonant scattering in the
\feka\ line and the measured Fe abundance is reliable. This result
also argues against delayed detonation models for SN Ia
and illustrates the power of X-ray observations of clusters of
galaxies to learn about supernova physics. We note that there are no
measurements of Ni abundances from observations of Galactic supernova
remnants because these objects are too cool to provide significant 
counts in the Ni K lines.

\section{
Perseus
                               }

The Perseus cluster is one of the closest, X-ray bright, rich cluster
of galaxies.  Its X-ray emission, has been the object of many studies
since its discovery as an X-ray source by Fritz et al. (1971). It is a
nearby cluster (at a redshift of 0.0183) and Bautz-Morgan type
II-III. The cluster is elongated and the ratio of its minor to major
axis is 0.83 at radii greater than 20' (Snyder et al. 1990). The
radiative cooling time of the X-ray emitting gas in the central
regions of Perseus is less than a Hubble time so the cluster has a
cooling flow with a mass deposition rate of about (1--5)$\times 10^{2}$
 M$_{\odot}$ yr$^{-1}$ (Fabian et al. 1981, Allen et al. 1992; Peres et al. 1998).  
 The centroid of the
cluster emission is offset by $\sim$ 2$^\prime$ to the east of NGC 1275 (Snyder et
al. 1990, Branduardi-Raymond et al. 1981), which has a Seyfert nucleus
and shows signs of non-thermal emission with a power-law
spectrum (Rothschild et al. 1981, Primini et al. 1981,
Branduardi-Raymond et al. 1981, Ulmer et al. 1987, Kowalski et al. 1993, Allen et al. 1999). The average
temperature of the X-ray emitting gas is approximately 6.5 keV (Eyles
et al. 1991) and the average abundance is 0.27 Solar in the central 1
degree (Arnaud et al. 1994). There are significant signs of both
temperature and abundance substructure (Schwarz et al. 1992, Arnaud
et al. 1994) in the gas distribution.

The existence of an iron abundance gradient in Perseus was first
suggested by Ulmer et al. (1987) using data from SPARTAN 1. They found
that the inner (0-5$^\prime$) region has an abundance of $\sim$ 0.81
Solar and temperature $\sim$ 4.2 keV while an outer region (6-20$^\prime$) 
has an abundance of $\sim$ 0.41 Solar and a temperature of
$\sim$ 7.1 keV. Further observations of Perseus have confirmed the
gradient. Ponman et al. (1990) analyzing data from Spacelab 2 found an
abundance value of $\sim$ 0.75 Solar in the central 3$^\prime$ which
decreases to $\sim$ 0.25 Solar in the outer regions. Kowalski et
al. (1993) also using SPARTAN 1 data have measured a central abundance
enhancement with a central value of $\sim$ 0.77 Solar. {\sl ASCA} analysis
of GIS data (Arnaud et al. 1994, see also Fabian
et al. 1994 for SIS data from the central region) has
corroborated the existence of a global abundance gradient (with
local non-axisymmetric variations), where the abundance values ranged 
from $\sim$ 0.4 Solar in the central regions to 0.2 Solar in the outer 
regions, over a region of $\sim$ 1 degree around the cluster center.

%===============================================================================
                                \section{
Data Reduction, Temperature \& Abundance Profiles
                                }

{\sl ASCA} carries four large-area X-ray telescopes, each with its own
detector: two Gas Imaging Spectrometers (GIS) and two Solid-State
Imaging Spectrometers (SIS).  Each GIS has a 50$^\prime$ diameter circular
field of view and a usable energy range of 0.8--10 keV; each SIS has a
22$^\prime$ square field of view and a usable energy range of 0.4--10 keV.
 
Two pointings of the Perseus cluster were analyzed in this
work. Perseus was observed for 20 ksec by {\sl ASCA} in August of
1993 (central pointing) and in September of 1993 (outer pointing) .
We selected data taken with high and medium bit rates, with cosmic ray
rigidity values $\ge$ 6 GeV/c, with elevation angles from the bright
Earth of $\ge20^{\circ}$, and from the Earth's limb of $\ge5^{\circ}$
(GIS) or $10^{\circ}$ (SIS); we also excluded times when the satellite
was affected by the South Atlantic Anomaly.  Rise time rejection of
particle events was performed on GIS data, and hot and flickering
pixels were removed from SIS data.  The resulting effective exposure
times for each instrument are shown in Table 1.  We estimated the
background from blank sky files provided by the {\sl ASCA} Guest
Observer Facility.
%----------------------------------------
{\sl ASCA} PSF mostly scatters high 
energy photons from internal regions outwards. The scattering from 
outside inwards is negligible and should not affect our results. The two  
extraction regions chosen for the central pointing ($0-4^\prime$ and $0-20^\prime$) 
are centered on the 
contaminating source, i.e., the X-ray center. The gas temperature in the very 
central region ($0-4^\prime$) is relatively 
low ($\approx$ 4.5 keV). Therefore, the effect of high energy 
photon depletion in that region due to the PSF is negligible 
(Takahashi et al. 1995 
\footnote{http://heasarc.gsfc.nasa.gov/Images/asca/newsletters/ext\_src\_analysis3.gif}). 
Furthermore, the best-fit temperature for that region agrees very well
with that of White (2000), who did correct for the PSF, when the same  
fitting spectral models are used (see below). Therefore, we do not worry 
about {\sl ASCA} PSF scattering. 
We also include the analysis of a $0-20^\prime$ region centered in the X-ray peak, which 
is more than three times larger than that where Molendi et al. detected resonant scattering.
 We analyzed this larger region for the purpose of 1) measuring the elemental abundances
in a region large enough so that resonant scattering is not effective, 2) helping to discriminate 
between different delayed detonation SN Ia explosion models with better photon statistics and
3) giving a better idea of the uncertainties involved when a power law spectral component is introduced.
%----------------------------------------------------------------------

We used XSPEC v10.0 (Arnaud 1996) to analyze the SIS and GIS spectra
separately and jointly. The spectra were fit using the {\tt mekal}
and {\tt vmekal} thermal emission models, which are based on the
emissivity calculations of Mewe \& Kaastra (cf. Mewe, Gronenschild \&
van den Oord 1985; Mewe, Lemen \& van den Oord 1986; Kaastra 1992),
with Fe L calculations by Liedahl, Osterheld \& Goldstein (1995).
Abundances were measured relative to the Solar photospheric values of
Anders \& Grevesse (1989), in which Fe/H=$4.68\times10^{-5}$ by
number.  Galactic photoelectric absorption was incorporated using the
{\tt wabs} model (Morrison \& McCammon 1983). Spectral channels were
grouped to have at least 25 counts/channel. Energy ranges were
restricted to 0.8--10 keV for the GIS and 0.4--10 keV for the SIS.
We show here the results of the spectral analysis  
for three spatial regions. The first is a circular region of 4$^\prime$ radius around the cluster's
X-ray center,
chosen for comparison with equivalent spatial regions in A496, A2199 
and A3571. The second is a region of 10$^\prime$ radius centered 35$^\prime$
from the center (outer pointing), selected to provide a measurement of abundances well
outside the core. The third is a circular region of 20$^\prime$ radius around the cluster's
X-ray center (basically covering the whole effective field of view of the central pointing).

Since there is a moderate cooling flow in the center of the Perseus
cluster we added a cooling flow component to the {\tt mekal} isothermal
emission model in the central region.  We adopted the emission measure
temperature distribution that corresponds to an isobaric cooling
flow. We tied the maximum temperature of the cooling flow to the
temperature of the isothermal component, and we fixed the minimum
temperature at 0.1 keV.  The abundances of the two emission components
({\tt mekal} and {\tt cflow}) were tied together.  We also applied a
single (but variable) Galactic absorption to both spectral components {\tt wabs}
and a variable extra absorption {\tt zwabs} to the cooling flow component.

We fitted spectra from the two SISs and two GISs independently and then from all
four instruments together. The spectral fits of SIS0 data for the
central pointing had unusually high reduced chi-squared ($\chi_{\nu}^{2}$)
 and showed
spurious spectral features. In particular, the width of the Fe K complex
is 50\% larger in SIS0 than SIS1. We have no explanation for this 
discrepancy. Nevertheless, the SIS0 best-fits for the interesting parameters 
agreed well with those obtained with SIS1 and the two GISs. Since the 
inclusion of the SIS0 data did not alter the results significantly and 
made the $\chi^{2}$ significantly worse we excluded the instrument 
from our analysis. 

When performing joint fits, the redshifts parameters correspondent to spectral models applied to 
GISs and SIS1 were let free to vary to compensate for possible gain differences between different instruments.
 The individual and joint SIS1 and GIS fits achieved $\chi^{2}_{\nu}$
 close to 1 and produced consistent estimates of temperatures
and metal abundances.  
The best-fit values for the temperatures and abundances for the central 
4$^\prime$ and for the external pointing are shown in
Table 2.  It can be seen from Table 2 that there is a positive temperature gradient. The best-fit temperature 
found for the outer region is $7.24 \pm 0.4$ keV and it declines to $4.43\pm 0.08$ keV in the inner 4$^\prime$, which
is consistent with the presence of a central cooling flow.  There is a significant ($> 90\%$ confidence) central abundance
enhancement. The abundance in the center is $0.50 \pm 0.03$ Solar
declining to $0.25 \pm 0.05$ Solar at $\sim$ 35$^\prime$ from the center. 

%===============================================================================

				\subsection{
 Individual Elemental Abundances
                                }

We determined the abundances of individual elements using a fitting
procedure similar to that used for determining general abundances, but with
the {\tt vmekal} spectral model in XSPEC. Since the global abundance is 
mainly driven by Fe, the cooling flow abundance was tied 
to the Fe abundance of the {\tt vmekal} component. We adopt as a basis for
comparison the following theoretical elemental abundance (by number) ratios
relative to Solar values : for SN Ia (Nomoto, Thielemann
\& Yokoi 1984, Nomoto et al. 1997a),
$$
{\rm O} \approx  {\rm Mg} \approx  0.035  {\rm Fe},
$$
$$
{\rm Ne} \approx  0.006 {\rm Fe},
$$
$$
{\rm Si} \approx  {\rm S} \approx  {\rm Ar} \approx  {\rm Ca} \approx 0.5 
{\rm Fe},
$$
$$
{\rm Ni} \approx  4.8 {\rm Fe},
$$
and for  SN  II (Nomoto et al.\ 1997b), 
$$
{\rm O} \approx  {\rm Mg} \approx  {\rm Si} \approx 3.7 {\rm Fe},
$$
$$
{\rm Ne} \approx {\rm S} \approx  2.5 {\rm Fe},
$$
$$
{\rm Ar} \approx  {\rm Ca} \approx  {\rm Ni} \approx 1.7 {\rm Fe}.
$$

In our
spectral model fits, the He abundance was fixed at the Solar value,
while C and N were fixed at 0.3 Solar (the derived abundances of other
elements are not affected by the particular choice of C and N
abundances).
%------------
In order to test the statistical significance of the spectral fittings
where individual elemental abundances are free to vary we used the F-test.
Initially, all the individual abundances were tied. Then we systematically 
untied the abundance of each interesting element, refitted and a new $\chi^{2}$ 
was found. The difference between the $\chi^2$ of these two fits must follow 
a $\chi^2$ distribution with one degree of freedom (Bevington 1969). Based
on the $\chi^2$ differences for each spectral fittings the F-test 
shows significant differences at the $>$99.5\%, $>$99.9\%, $>$99\%, $>$99.9\%,
$>$97.5\% 
and $>$95\% confidence for letting Fe, Ni, Si, Ne, O, and S free to vary, 
respectively. The significance is $<$ 90\%  for 
Mg, Ar and Ca, and therefore they are tied together.
We used in our
analysis only those abundances which were reasonably well constrained,
namely O, Ne, Si, S, Fe, \& Ni.  The observed individual abundances are shown in
Table 3 for the central $0-4^\prime$ and for an outer circular region centered at $\sim
35^\prime$ away from the center with a radius of 10$^\prime$. The
individual elemental abundances in the outer parts were not 
constrained enough to determine spatial variations of abundance ratios within
the cluster.  We, therefore, could only set upper limits for most 
individual elemental abundances other than Fe in the outer region.

We also determined the central Fe abundance from the Fe~L lines alone.
The abundance found when the Fe~K lines are ignored is 0.62
$^{+0.14}_{-0.12}$ and is also shown in Table 3.  This value is
consistent with the abundance determined from the Fe~K complex. The 
Ni abundance determined from the L shell lines is not well constrained 
(1.3$^{+1.4}_{-1.1}$) but it is consistent with that determined from the 
K shell complex\footnote{Ni abundances based on L shell lines are
likely to be unreliable for the following two reasons. Firstly, the 
Ni~L lines are interspersed among the Fe~L lines so an accurate 
measurement of the Ni~L line fluxes requires a good model for the Fe~L 
lines. Secondly, even if the Ni~L line fluxes are correct then
turning these into a Ni abundance depends on accurate Ni L shell
atomic physics. The corrections to the Fe L shell physics of Liedahl
\etal (1995) have not yet been applied to Ni so any conclusions based
on Ni~L lines should be considered very preliminary.}.
The best constrained individual abundances are Fe, Si,
and Ni. We also consider abundances of other elements when
determining the SN Ia Fe mass fraction in the next section.

%===============================================================================
				\section{
  Abundance Ratios and the SN Ia and SN II Fractions
                                }

Discriminating between the products of SNe Ia and II requires the
determination of the abundances and abundance ratios of
$\alpha$-elements and comparison to SN models.  The determination of
the relative contributions of the two SN types is complicated by the
theoretical uncertainties in the elemental mass yields for different
SN models, especially for SNe II (e.g. Gibson, Loewenstein \&
Mushotzky 1997). In general, SN Ia models show far better agreement on
elemental mass yields than SN II models. However, discrepancies
between some SN Ia explosion models still exist. For example, the Ni
mass yields for models with fast flame-speed like W7 (Nomoto,
Thieleman, \& Yokoi 1984; Thieleman, Nomoto \& Yokoi 1986), are $\sim
3$ times bigger than those for slow flame-speed (delayed
detonation). In view of these theoretical uncertainties, it is better
to use a combination of several different elemental abundance ratios when estimating
the SN Ia iron mass fraction, rather than basing analysis on a single
specific ratio.

To determine the SN Ia Fe mass fraction contribution we considered
several abundance ratios involving the elements described in the
previous paragraph.  The abundance ratios used are shown in Table 4,
along with the expected theoretical abundance ratios (by number,
normalized by the Solar photospheric values) for pure SN Ia (for both W7
and delayed detonation explosion model WDD2 of Nomoto et al. 1997a) and
SN II ejecta. All the errors associated with the observed abundance ratios
are the propagated 90\% confidence errors. The derived SN Ia Fe mass fractions
for the inner region ($0-4^\prime$) are also listed in Table 4.  It can be
seen that most abundance ratios show, within the 90$\%$ confidence
errors, a combination of SN Ia and SN II ejecta. Some abundance ratios, however, 
are observed to be completely out of the predicted theoretical range for SN Ia and
II, e.g. Ne/S \& Ne/Si. These measurements are consistent with the results of
Mushotzky \etal\ (1996) who also found that S is observed to be
underabundant and Ne overabundant relative to several different
theoretical SN II models. Based on a detailed analysis of A496, Dupke \& White
(2000b) proposed correcting the theoretical SN II sulfur yields by
0.25--0.5. To be conservative, we avoid including abundance ratios
involving these two elements in the calculation of the SN Ia iron mass
fraction.

From the abundance ratios involving O, Si, and Fe, we derived the fraction of
the iron mass from SN Ia. The Fe mass fraction estimates are also listed
in Table 4. The average SN Ia Fe mass fraction based on ratios involving the above mentioned
three elements is 0.69$\pm{0.08}$. This value is consistent with the
SN Ia Fe mass fraction derived from the Ni/Fe abundance ratio alone
(0.60$\pm{0.35}$). The total weighted average of the SN Ia Fe mass fractions
based on ratios involving only the abundances of Ni, O, Si, \& Fe is 0.67$\pm{0.06}$.
 The observed Ni/Fe ratio is inconsistent with either : 1) pure SN II ejecta
or 2) a combination of SN II and SN Ia ejecta {\it as predicted by delayed
detonation models}.  The mutual consistency of the derived SN Ia Fe
mass fractions derived from different abundance ratios suggests that
the Ni/Fe ratio is a robust indicator of SN Ia ejecta contamination.

%-------------------------------------------------------------------
It should be noticed that the Si/Fe ratio measured in the center (0-4$^\prime$) of Perseus 
(1.37$\pm{0.28}$) is completely out of the range allowed by the delayed detonation
model WDD1 (Si/Fe $>$ 1.69). Furthermore, if WDD2 is assumed there is a 
contradiction between the predicted SN Ia Fe mass fractions derived from 
Si/Fe (0.86$\pm{0.1}$) and O/Si ($<$0.73). However, the only abundance ratio 
that is systematically in contradiction with all three delayed detonation models is Ni/Fe.
%-------------------------------------------------------------------
Figure 1 shows that the Ni/Fe abundance ratio derived in this work is consistent
with the values found for other clusters that show central SN
Ia contamination. Figure 1 also shows the theoretical predictions
of the Ni/Fe ratio for SN II and for three different SN Ia delayed detonation
models. It is clear from the Figure that delayed detonation explosion models are not
preferred over the standard models such as W7. 

In order to have the SN Ia iron mass fractions derived from abundance
ratios involving Ne \& S to agree with those involving other elements
($\sim 0.67$), one would have to systematically multiply the theoretical SN II mass yields
of S by $\sim$ 0.34 and of Ne by $\sim$2.7.  These suggested corrections 
for Ne and S are also listed in Table 4.

%===============================================================================

\subsection{
Comparison with BeppoSAX results
                                }

Molendi et al. (1998) reported the detection of resonant scattering
in the central regions of the Perseus cluster based on the analysis of 
the BeppoSAX Medium Energy Concentrator
Spectrometers (MECS) data. In order to compare
their results with ours we have reanalyzed the BeppoSAX data for
Perseus. We used standard reduction procedures as recommended by the
BeppoSAX Science Data Center (www.sdc.asi.it).  The effective area
files were generated using ``EFFAREA'', which convolves a given
surface brightness profile with the MECS energy/position dependent PSF
and vignetting functions.

Following Molendi et al. we analyzed spectra from all three MECS using only
the channels corresponding to the 3--10 keV energy range. Background
spectra were extracted from blank-sky event files using
the same extraction region as that for the source. The spectral fitting
procedure was analogous to that used with {\sl ASCA} as described in the
third section of this paper.  We were unable to find reasonably good fits
with the redshift constrained to the optical value. This is probably
due to a known MECS gain calibration problem and the recommended
solution is to allow the redshift to be a free parameter (Fiore,
personal communication). We found a best-fit redshift of 0.011
\footnote{This gain problem has been observed to be significant in
80\% of the clusters observed by BeppoSAX and gives a redshift
discrepancy of $\approx 0.009$ when compared to optical redshifts
(J. Irwin, personal communication).}.

Because of the limited safe energy range of the MECS (3--10 keV) the
comparison of line strengths between BeppoSAX and {\sl ASCA} is limited to
the best constrained Fe and Ni lines. To compare with {\sl ASCA} we
extracted spectra from the 0-4$^\prime$ region. The results are
shown in Table 3. The reduced $\chi^2$ is significantly
worse than that obtained with {\sl ASCA} ($\chi_{\nu}^2 \approx 1.25$, 
for a {\tt vmekal} + {\tt cflow} model) even with the
cluster redshift set as a free parameter in the spectral fittings.  

%-----------------------------------------------------------------
The discrepancy of our best-fit values of Fe and Ni abundances
with those of Molendi et al. is most likely due to the difference in
dealing with the MECS gain calibration problem and to the insensitivity 
of the MECS spectra to the cooling flow within the energy range 
considered (3-10 keV). If we add a cooling flow component with the best-fit
normalization (300 M$_{\odot}$ yr$^{-1}$) obtained from the {\sl ASCA} spectral 
fittings, the best-fit Ni and Fe 
abundances are found to be 1.47$\pm0.48$  and 0.49$\pm0.01$,
respectively, with $\chi^{2}_{\nu}$ of $\sim$ 1.19.
%-----------------------------------------------------------------

We also extracted spectra
from a central region with a radius of 6$^\prime$ for the purpose of
direct comparison with the Molendi et al. best-fit values. Our
results are similar to those obtained for the 0-4$^\prime$ region. The
Ni abundance is 1.34$^{+0.26}_{-0.44}$ Solar and the Fe abundance is
0.47$\pm 0.012$ Solar with a $\chi_{\nu}^2 \approx 1.23$, for
a {\tt vmekal} + {\tt cflow} model. As can be seen in Table 3 the Fe and Ni abundance 
measurements in the central regions obtained with {\sl ASCA} and {\sl SAX} for 
the same spectral models agree very well. 

\subsection{
Influence of the Power Law Component
                                }

Since the cD galaxy in Perseus, NGC 1275, has a Seyfert-like spectrum
(Rubin et al. 1977), we added a power-law component to
the absorbed {\tt vmekal} + {\tt cflow} model. The addition of the power law 
component to the simultaneous GISs and SIS 1 fittings has a general effect 
of decreasing the temperature of the hot component (of the {\tt cflow} model)
 and does not improve 
the spectral fittings ($\chi^{2}_{\nu} \sim 1.12$ or $\chi^{2} \sim$ 1495 for 1333
degrees of freedom). However, the inclusion of the power law component introduces large
uncertainties in the determination of Ni \& Fe abundances and, especially, the power law
slope, which are found to be 1.5$\pm$0.7 Solar, 0.47$\pm$0.05 Solar and 2.6$^{+2.6}_{-0.7}$, respectively. 
In order to understand the reasons for the high uncertainties on these parameters introduced 
by the power law component, we fitted the spectra obtained with GISs and SIS 1 separately.
The individual spectral fittings indicated that the GIS 2\&3 are insensitive
to the inclusion of the power law. Furthermore, the introduction of the power
law component increases the uncertainties of the Ni abundance determination 
very strongly (90\% confidence range for Ni is 0.1-2.8). The SISs, on the other hand, are less sensitive to 
the {\tt cflow} model than to the power law component. There is a significant improvement
in the spectral fittings with the addition of the power law component. However, the Ni 
abundance ($\sim$ 1.6) as well as the other individual abundances are similar to the 
those obtained without the power law model. 

The allowed low Ni abundance is a consequence of the inclusion of the 
power law component in the joint spectral fittings and seems to be due
to different instrumental sensitivities in the spectral fittings to the combination
 {\tt cflow} + {\tt power} models, i.e., an artifact of the spectral fittings, 
 which also generates large uncertainties
in the best-fit values of the power law slope (varying from 0.9 to 3).
Since the introduction of the power law component does not improve significantly the 
simultaneous GISs + SIS 1 spectral fittings and
introduces artificial large uncertainties on the determination of the best-fit Ni 
abundances we do not include the power law component when determining the SN Ia Fe 
mass fraction.  However, since the power law component poses a new undesired uncertainty
to both mechanisms (resonant scattering and W7 SN Ia enrichment) proposed
to explain the excess flux at $\sim$ 8 keV, in the cluster central regions, we will 
consider its possible effects on the Ni and Ni/Fe measurements when 
checking the validity of the delayed detonation models (below).

In order to compare the two above mentioned competing mechanisms and to estimate the effect that a power 
law component would have in the abundance ratios involving Ni we extracted spectra from a large
region (radius of 20$^\prime$) around NGC 1275 in the central pointing. The size of the 
region is large enough so that the directional number of \feka\ photons is conserved and 
resonant scattering should have no effect 
on the \fekba\ ratio. The abundance ratios derived
from this region, however, are emission measured weighted averages and are highly
biased towards the values of the abundance ratios in the cluster central regions. 
The best-fit values of Si, Fe, Ni and their ratios obtained from the spectral 
fittings of the 0-20$^\prime$ region are shown in Table 5. 
The spectral fittings are significantly worse ($\chi^{2}_{\nu} \sim 1.79$) than 
those for the central 4$^\prime$. The observed Ni/Fe ratio (3.9$^{+0.6}_{-0.7}$) corresponds 
 to a SN Ia Fe mass fraction of 
0.73$^{+0.21}_{-0.23}$, which is consistent with that derived from the central 
4$^\prime$. This high value for the Ni/Fe ratio is inconsistent with all
delayed detonation models, independently of resonant scattering effects. The inclusion of
a power law component (with a best-fit slope of $\sim 1.2$) improves significantly the spectral fittings
($\chi^{2}_{\nu} \sim 1.55$) and has the effect of lowering somewhat the Ni abundance and,
therefore, the Ni/Fe ratio to 2.9$^{+0.6}_{-0.7}$, which is consistent with 
a radial decrease in SN Ia Fe mass fraction (Figure 1). However, even the lower Ni/Fe ratio
obtained with the addition of the power law component is in contradiction 
with the delayed detonation models. If the power law slope is fixed at higher values
 the Ni/Fe ratios measured are even higher. So, {\it the average Ni/Fe ratio is inconsistent
with the delayed detonation SN Ia explosion models independently of the resonant scattering
effect and also of the inclusion of a power law component in the spectral fittings.}

                           \section{
Discussion
}

\subsection{Abundance and Chemical Gradients}

The comparative spectral analysis of the large region versus the 
central region clarifies the question of whether a chemical gradient exists in the Perseus cluster.
The average (0-20$^\prime$) Si/Fe ratio measured is 1.83$^{+0.27}_{-0.10}$ (implying an emission
measure weighted average SN Ia Fe mass fraction of 0.57$^{+0.03}_{-0.09}$). This value is
significantly ($\ge$90\% confidence) higher 
than that obtained in the central regions (Si/Fe $=$ 1.37$\pm0.28$), indicating
 {\it the presence of a chemical gradient in the intracluster gas of Perseus, where 
SN Ia ejecta is more concentrated towards the central regions of the cluster.}
Some of the possible scenarios,
proposed to explain the abundance and chemical gradient mentioned before 
could be better constrained by determining the abundance radial 
profiles for different elements and analyzing the SN Ia and II Fe
mass-to-light ratio distribution for the cluster. This is beyond
the scope of this work.  Nevertheless, at least two difficulties are
apparent if stripping models are used to explain the central abundance and chemical 
gradients (Dupke \& White 2000b). Firstly, the Fe abundances measured in most early-type
galaxies by {\sl ASCA} (Loewenstein et al.\ 1994; Matsumoto et al.\
1997) and {\sl ROSAT} (Davis \& White 1996) are lower (0.2--0.4 Solar)
than the abundance observed at the cluster center ($\sim 0.5$
Solar)\footnote{However, there is evidence that the Fe abundance in
ellipticals can be significantly higher when multiphase spectral
models are used in the X-ray spectral fittings (Buote \& Fabian 1998).}.
Furthermore, there is evidence of abundance gradients in
elliptical galaxies themselves (e.g. in NGC 4636, Matsushita et al.\
1997), so that most stripped gas will have even lower abundances than
indicated by the global X-ray Fe abundances of elliptical galaxies. 
Secondly, the efficiency of ram-pressure stripping 
is dependent on the ICM density, which rises strongly towards the central regions. 
The difference in Fe abundances measured in inner and outer regions is
mild (a factor of $\sim 2$) and might be expected to be larger if
stripping were important. 

\subsection{
Resonance Scattering or SN Ia Convective Deflagration Models?
}

We have shown above that the high \fekb\ to \feka\ ratio observed in 
Perseus, which has been attributed to resonant scattering of the \feka,
can also be explained by a supersolar Ni/Fe abundance ratio due to
the products of SNe Ia. In this section we
consider the arguments for and against these two explanations. The
first point to make is that this is independent of the question of
whether there is a significant contribution from SN Ia products. The
abundance ratios of elements other than Ni can be used to measure
the relative contributions from SN Ia and SN II. 
So, the basic choice lies between a) resonance scattering and delayed 
detonation SN Ia explosion models or b) no resonance scattering and 
convective deflagration SN Ia explosion model.

Delayed detonation SN Ia explosion models (and alternatives which have low Ni production) have certainly
been espoused by SN theorists. However, recent treatment of turbulence effects in 
subsonic flames in delayed detonation models shows that the transition to detonation 
requires unusually high turbulent velocity fluctuations (Lisewski, Hillebrandt \& Weaver 2000). 
%--------------------------------------------------------------------
Resonant scattering optical depth is a function of the gas density and, therefore, should 
fall to the optical thin value at regions away from the cluster's center. Observations of the
radial distribution of the SN Ia gas contamination in other clusters also show a decrease in SN Ia
Fe mass fraction as the distance from the center increases (Dupke 1998, Dupke \& White 2000b, 
Finoguenov, David, \& Ponman 2000). This seems also to be the case for Perseus as indicated 
by significantly lower Si/Fe value in the central 4$^\prime$ when 
compared to the average value observed for the whole cluster (0-20$^\prime$).
 This ``chemical gradient'' is 
expected from models proposed to explain the central SN Ia 
ejecta excess (e.g. suppressed SN Ia winds, ram-pressure stripping, cD stellar mass loss). Therefore, 
if one assumes the classical W7 SN Ia model, both scenarios, SN Ia enrichment and resonant scattering,
predict that, observationally, the excess flux in the \fekb\--\nika\ line complex should decrease radially.

Against the resonance scattering scenario is the fact that simulations of its expected 
effect on clusters (where Ni abundances are set to Solar) show that resonant scattering 
alone cannot explain the high \fekb\ to \feka\ ratio observed  (Tawara et al. 1997, Furuzawa
et al. 1997) and, in particular, in Perseus (Akimoto et al. 1997, 1999)\footnote{Actually, the intensity ratio 
$\frac{FeK\alpha}{(FeK\beta + NiK\alpha)}$ for 
the central regions of 78 clusters (Figure 1 of Akimoto et al. 1999) can be well described 
by an optically thin plasma with a central Ni/Fe ratio of $>$ 2. }.
%---------------------------------------------------------------------------
Furthermore, if the X-ray emitting gas has turbulent velocities of a few hundred km/s then any 
resonance scattering will be totally suppressed (Gilfanov et al. 1987). 

If we use Fe, Si, and O alone to
set the relative fractions of SN II and SN Ia products and then use these relative fractions 
to predict Ni relative abundances we obtain a very good match to what is observed if the W7 model 
for SN Ia is assumed. Therefore, resonant scattering can  contribute significantly 
to the relative enhancement of the \fekb\--\nika\ line complex only if delayed detonations are 
assumed. Chemically W7 and WDD models differ in the ejected mass yields of some elements. In
particular WDDs produce more Ni than W7 and, in a smaller degree, more Si. If resonant scattering 
were the mechanism responsible for the apparent high Ni abundances in the central cluster regions, and, 
therefore, WDD models were preferred, then we should expect that if one extracted spectrum from 
a region large enough, such as to be immune to resonant scattering, the observed Ni/Fe abundance 
ratio would be smaller and, necessarily, consistent with WDD models. In the previous paragraph we showed 
that the Ni/Fe ratio measured within the 0-20$^\prime$ region is not consistent with any WDD model, even 
when the uncertainties from the power law component are incorporated. This shows that delayed 
detonation models are not preferred over the W7 model and strongly 
suggests that resonant scattering has a much smaller contribution (if any) to the observed 
Fe abundance measurements than that claimed by Molendi et al.

Molendi et
al. determination of the resonant scattering in Perseus was based on
spectral fittings of a blend of Fe and Ni lines, assuming the contribution of the
\nika\ line to this blend to be small, therefore, requiring {\it a
priori}, a specific SN enrichment Type (either SN II or delayed
detonation SN Ia). One of the goals of our analysis is to
determine the very SN Type contamination of the ICM. We showed
that the delayed detonation SN Ia explosion models are not adequate to explain 
the observations while the ``classical'' W7 SN Ia model can explain what we observe
without the need to invoke resonant scattering. 
 
%===============================================================================
                           \section{
Conclusions
}

In this work we have shown that 
\begin{enumerate}
\item The central region (0-4$^\prime$) of the Perseus cluster is significantly
enriched with ejecta from SN Ia, which produce more than half ($\sim
67\%$) of the total iron.
\item The observed Ni/Fe ratio is in excellent agreement with the observed ratios
of other elements involving Fe, Si, Ni and O in predicting the SN Ia Fe mass fraction in the 
central regions of Perseus . The Ni/Fe ratio observed is $\sim$ 3.4 .
\item The Ni/Fe ratio is consistent with the theoretically predicted values
for SN Ia models with fast flame-speed (W7) but not with delayed
detonation explosion models.
\item The average Si/Fe ratio in Perseus is significantly ($>$90\% confidence)
higher than in the central regions, which indicates that the central Fe abundance
gradient is accompanied by a chemical gradient, i.e., the SN Ia contamination 
has a radial decline in Perseus. 
\item It is not necessary to invoke resonant scattering in the \feka\ 
line since the high \fekb/\feka\ ratio can be totally attributed to
substantial ICM contamination from SN Ia ejecta.  
\end{enumerate}

Although the high central Ni abundances can be fully explained by SN Ia 
enrichment, the current data do not allow a precise quantification of 
the relative contribution of resonant scattering, which can be done 
only with very high resolution X-ray spectroscopy.  
 If resonance scattering is important it will always
provide the largest effect at the peak density of the
cluster. Therefore, clusters with high resonant scattering
optical depth should show, observationally, a Ni abundance profile decreasing with
radius (within the resolution of current X-ray spectrometers). 
Therefore, additional discriminating
evidence between the two scenarios can be provided by observation of
other clusters that show no central SN Ia ejecta dominance and, at the same time, have large resonant 
scattering optical depth.

\acknowledgments RAD thanks the USRA for the
talk invitation that originated this work and NASA Grant NAG 5-3247 
for partial support. We thank Raymond White III,
Fumie Akimoto, Melville Ulmer and Barbara Eckstein for helpful
suggestions.  We would also like to thank F. Fiore for helpful
suggestions.  RAD would like to thank Jimmy Irwin for very helpful
discussions and for checking the observed BeppoSax calibration
problems for 10 other clusters and Phillipe Fischer for helpful
insights and suggestions. This research made use of the HEASARC {\sl
ASCA} database and NED.

\clearpage

%===============================================================================
				
                                \clearpage

%===============================================================================
                                \begin{figure}
                                \title{
Figure Captions
                                }

                                \caption{
Ni/Fe abundance ratio measurements (by number normalized to Solar) of A496, A2199, 
A3571 and Perseus (A426) inner regions ($\sim 125 h_{50}^{-1}$ kpc). It is also shown the 
theoretical predictions for pure SN II enrichment (dotted line) and different models of pure SN Ia 
enrichment: the ``Convective Deflagration'' W7 (thick line) and Delayed Detonation models 1,2,3 of 
Nomoto et al. 1997a (thin lines). For Perseus we also show the Ni/Fe abundance ratio from 
spectral fittings of a region encompassing a region of 20$^\prime$ radius with 
(PL) and without (noPL) an extra power law component.
                                }
                                 
                                \end{figure}
\clearpage
%=== Table 1 ===================================================================
\begin{deluxetable}{lccc}
\small
\tablewidth{0pt}
\tablecaption{Effective Exposure Times}
\tablehead{
\colhead{Spectrometer} &
\multicolumn{2}{c} {Exposure Time (ksec)} & \nl
\colhead{} &
\colhead{Central Pointing}  &
\colhead{Outer Pointing}  &
}
\startdata
SIS 0 & 18 & 18 \nl
SIS 1 & 18 & 19 \nl
GIS 2 & 12 & 17 \nl
GIS 3 & 12 & 17 \nl
\enddata
\end{deluxetable}
\clearpage
%=== Table 2 ===================================================================
\begin{deluxetable}{lcccc}
\small
\tablewidth{0pt}
\tablecaption{Spectral Fits\tablenotemark{a}}
\tablehead{
\colhead{Region} &
\colhead{kT} &
\colhead{Abundance} & 
\colhead{$\chi_{\nu}^{2}$} \nl
\colhead{} &
\colhead{(keV)}  &
\colhead{(Solar)}  & 
\colhead{} &
}
\startdata
$0 - 4^\prime$  & 4.43$^{+0.08}_{-0.08}$ & 0.50$^{+0.030}_{-0.025}$ & 1.29 \nl
$25 - 45^\prime$  & 7.24$^{+0.44}_{-0.40}$ & 0.25$^{+0.05}_{-0.05}$ & 1.05\nl
\enddata
\tablenotetext{a}{Errors are 90\% confidence limits}
\end{deluxetable}
\clearpage
%=== Table 3 =====================================================================
\begin{deluxetable}{lccc}
\small
\tablewidth{0pt}
\tablecaption{Individual Elemental Abundances for the Central and Outer Regions of Perseus \tablenotemark{a,b}}
\tablehead{
\colhead{Element} &
\colhead{Abundance (inner)} &
\colhead{Abundance (outer)} &\nl
\colhead{} &
\colhead{(Solar)} &
\colhead{(Solar)} & \nl
}
\startdata
O & 0.87$^{+0.63}_{-0.53}$ & $\le 0.95$ &\nl
Ne & 1.19$^{+0.39}_{-0.41}$ & $\le 2.01$ &\nl
Si & 0.71$\pm 0.14$ & $\le 0.83$ &\nl
S & 0.32$\pm 0.16$ & $\le 1.11$ &\nl
Fe & 0.52$^{+0.030}_{-0.025}$ & 0.25$^{+0.05}_{-0.05}$ &\nl
Fe$_{L}$\tablenotemark{c} & 0.62$^{+0.14}_{-0.12}$ & &\nl
Fe$_{SAX}$\tablenotemark{d} & 0.50$^{+0.013}_{-0.012}$ & &\nl
Fe$_{SAX}$\tablenotemark{e} & 0.49$\pm 0.01$ & &\nl
Ni & 1.79$^{+0.57}_{-0.56}$ & $\le 1.5$ &\nl
Ni$_{SAX}$\tablenotemark{d} & 1.17$^{+0.29}_{-0.32}$ & &\nl
Ni$_{SAX}$\tablenotemark{e} & 1.47$^{+0.48}_{-0.49}$ & &\nl

\tablenotetext{a}{Errors are 90\% confidence level limits}
\tablenotetext{b}{Simultaneous {\sl ASCA} spectral fittings using GIS 2\&3 and SIS 1 spectrometers for an absorbed {\tt Vmekal} + {\tt Cflow} model ($\chi^{2}_{\nu}$ = 1.12)}
\tablenotetext{c}{Fe abundance determined from spectral fits of the FeL line complex}
\tablenotetext{d}{Abundance determined from simultaneous spectral fits of MECS 1, 2 \& 3, for a simple {\tt Vmekal} model}
\tablenotetext{e}{Abundance determined from simultaneous spectral fits of MECS 1, 2 \& 3, for a simple {\tt Vmekal} + {\tt Cflow} model}
\enddata
\end{deluxetable}
\clearpage
%=== Table 4 =====================================================================
\begin{deluxetable}{lcccccccc}
\small
\tablewidth{0pt}
\tablecaption{Elemental Abundance Ratios\tablenotemark{a}}
\tablehead{
\colhead{Abundance} &
\colhead{Region} &
\multicolumn{3}{c} {Theory\tablenotemark{b}} &
\colhead{SN Ia Fe Mass Fraction} &
\multicolumn{2}{c} {Correction\tablenotemark{c}} &\nl
\colhead{Ratio} &
\colhead{$0 - 4^\prime$ }  &
\colhead{SNIa$_{W7}$} &
\colhead{SNIa$_{WDD2}$} &
\colhead{SNII} &
\colhead{$0 - 4^\prime$} &
\colhead{S} &
\colhead{Ne} \nl
}
\startdata
O/Fe & 1.67$^{+0.76}_{-0.79}$ &  0.037 & 0.019 & 3.82 
& 0.57$\pm$0.21 & & & \nl
Si/Fe & 1.37$\pm$ 0.28 &   0.538 & 1.01 & 3.53 
& 0.72$\pm$0.09 & & & \nl
Si/Ni & 0.40$\pm$0.15 &  0.113 & 0.725 & 2.14 
& 0.68$^{+0.15}_{-0.12}$ & & & \nl
Ni/Fe & 3.44$\pm$ 1.1 &  4.758 & 1.40 & 1.65 
& 0.60$^{+0.35}_{-0.36}$ & & & \nl
Ni/Fe$_{L}$ & 2.89$^{+1.14}_{-1.06}$ & 4.758 & 1.40 & 1.65 
& 0.43$^{+0.37}_{-0.36}$ & & & \nl
O/Si & 1.23$\pm$ 0.6 &  0.068 & 0.019 & 1.1 
& $<$0.84 & & & \nl
O/Ni & 0.49$\pm$0.27 &  0.008 & 0.014 & 2.32 
& 0.57$^{+0.20}_{-0.15}$ & & & \nl
O/S & 2.72$^{+1.83}_{-1.87}$ & 0.092 & 0.084 & 1.54 
& $<$0.8 & 0.11 & & \nl
Si/S & 2.22$\pm$1.19 & 0.063 & 0.016 & 1.67 
& $<$0.95 & 0.39& & \nl
S/Fe & 0.62$\pm$0.31 &  0.585 & 1.2 & 2.29 
&0.98$^{+0.02}_{-0.18}$ & 0.3 & & \nl
S/Ni & 0.18$\pm$0.11 &  0.123 & 0.86 & 1.39 
&0.88$^{+0.12}_{-0.18}$ & 0.37 & & \nl
Ne/S & 3.72$^{+2.22}_{-2.26}$ & 0.011 & 0.001 & 1.18 
& ---& 0.34 & 2.7 & \nl
Ne/Fe & 2.29$^{+0.76}_{-0.80}$ &  0.006 & 0.001 & 2.69 
&0.15$^{+0.30}_{-0.15}$ & & 2.6& \nl
Ne/Si & 1.68$^{+0.64}_{-0.67}$ &  0.012 & 0.001 & 0.76 
& --- & & 2.9 & \nl
Ne/Ni & 0.67$\pm$0.30 &  0.001 & 0.001 & 1.63 
& 0.33$^{+0.21}_{-0.14}$ & & 2.8& \nl
\enddata
\tablenotetext{a}{Errors are  propagated 90\% errors}
\tablenotetext{b}{SN Ia: Nomoto et al (1997a); SN II: Nomoto et al (1997b)}
\tablenotetext{c}{Number that if multiplied by the SN II mass yields of Ne or S would bring 
the SN Ia mass fraction to the average defined by ratios involving Fe, Si, Ni \& O}
\end{deluxetable}
%=== Table 5 =====================================================================
\begin{deluxetable}{lcc}
\small
\tablewidth{0pt}
\tablecaption{Average Individual Elemental Abundances and Abundance Ratios for  Perseus \tablenotemark{a}}
\tablehead{
\colhead{Element} &
\colhead{Abundance or Abundance Ratio (Average)\tablenotemark{a} } &\nl
\colhead{} &
\colhead{(Solar)} & \nl
}
\startdata
Si & 0.88$^{+0.13}_{-0.04}$  &\nl
Fe & 0.48$^{+0.012}_{-0.015}$ &\nl
Fe$_{power}$\tablenotemark{b} & 0.5$^{+0.02}_{-0.02}$&\nl
Ni & 1.87$^{+0.27}_{-0.33}$ &\nl
Ni$_{power}$\tablenotemark{b} & 1.46$^{+0.34}_{-0.33}$ &\nl
Si/Fe & 1.83$^{+0.27}_{-0.10}$ &\nl
Ni/Fe & 3.9$^{+0.6}_{-0.7}$ &\nl
Ni/Fe$_{power}$\tablenotemark{b} & 2.92$\pm$0.7 &\nl
\enddata
\tablenotetext{a}{Errors are 90\% confidence level limits or propagated 90\% confidence level}
\tablenotetext{b}{Abundance determined with the addition of a power law model to {\tt Vmekal} + {\tt Cflow}}
\end{deluxetable}

\clearpage

                                
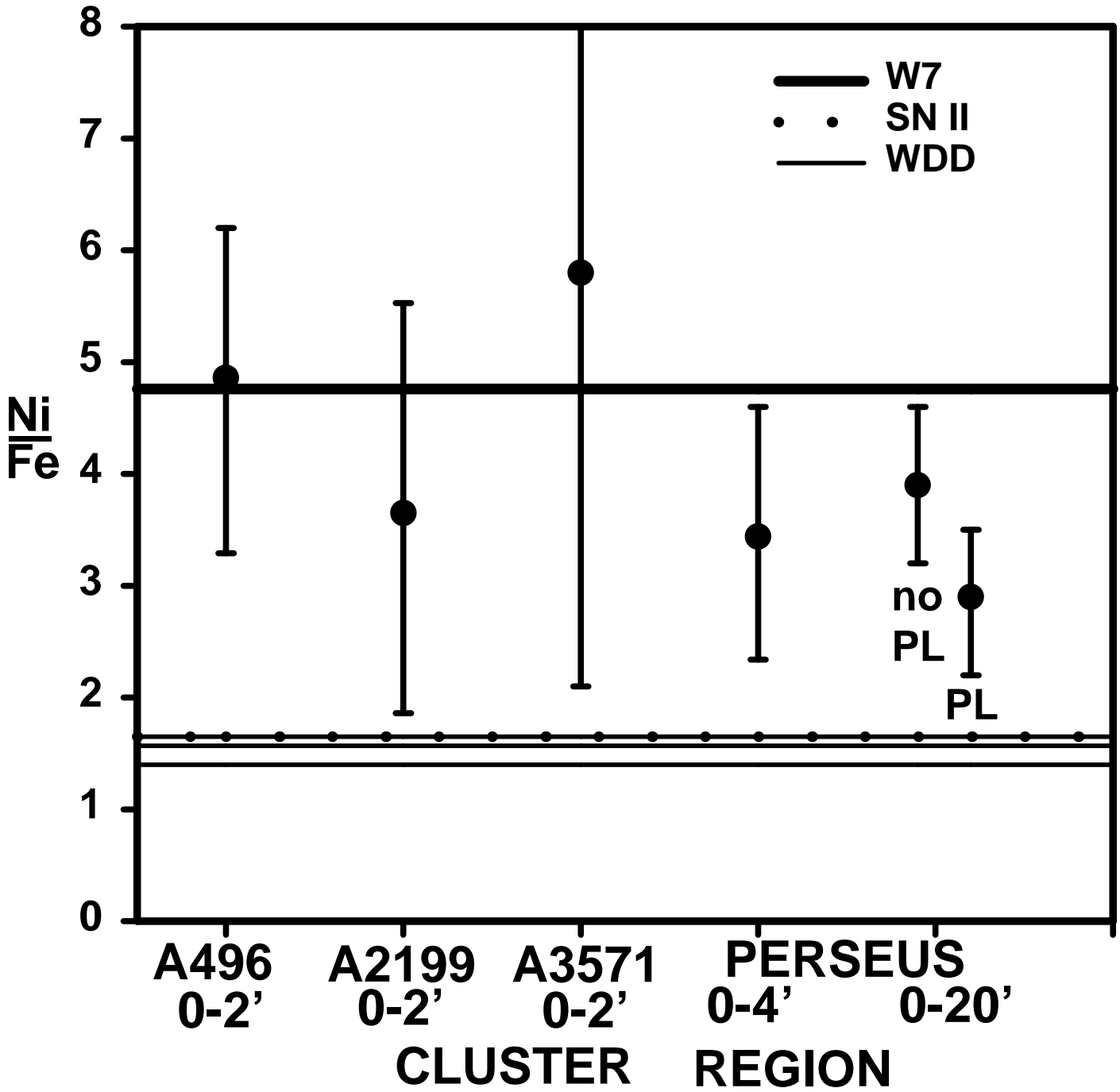
\end{document}